\documentclass[conference]{IEEEtran}
\usepackage{multirow}
\IEEEoverridecommandlockouts
\usepackage{url}
\usepackage{booktabs}
\usepackage{times} 
\usepackage{cite}
\usepackage{amsmath,amssymb,amsfonts}
\usepackage{algorithmic}
\usepackage{graphicx}
\usepackage{textcomp}
\usepackage{xcolor}
\usepackage{type1cm}
\def\BibTeX{{\rm B\kern-.05em{\sc i\kern-.025em b}\kern-.08em
    T\kern-.1667em\lower.7ex\hbox{E}\kern-.125emX}}
\begin{document}

\title{RUBICONe: Wireless RAFT-Unified Behaviors for Intervehicular Cooperative Operations and Negotiations}

\author{Zhenghua Hu$^{*}$, Tairan Dan$^{*}$, Zeyu Tao, Jiacheng Qian, \\
Amedeo Morat, Lorenzo Romano, Alessandro Massafra, and Hao Xu$^{\dagger}$, \IEEEmembership{Member,~IEEE}

\thanks{*Z. Hu and T. Dan contributed equally to this work.}
\thanks{$^{\dagger}$H. Xu is the corresponding author.}
\thanks{All authors are with Tongji University, Shanghai 201804, China. A. Morat, L. Romano, and A. Massafra are also with Politecnico di Torino, 10129 Turin, Italy. 
(E-mail: 2353741@tongji.edu.cn, 2453796@tongji.edu.cn, 2253879@tongji.edu.cn, 2230919@tongji.edu.cn, 2359267@tongji.edu.cn, s322011@studenti.polito.it, 2359275@tongji.edu.cn, hxu@tongji.edu.cn).}
}

\maketitle

\begin{abstract} Just as Caesar declared ``alea iacta est'' (the die is cast) upon crossing the Rubicone river, lane change decisions in autonomous vehicles also represent critical points of no return. RUBICONe addresses this challenge by recognizing that lane change decision-making relying solely on a single vehicle's perception would be as precarious as crossing an unknown river alone. By implementing a distributed consensus framework that extends the RAFT algorithm with wireless connectivity, RUBICONe enables multiple vehicles to collectively process and aggregate their perceptions. Using multiple software-defined radio (SDR) devices as the experimental platform, this study demonstrates how consensus-based decision-making significantly reduces the impact of environmental interference and mitigates the risk of misjudgments by individual vehicles. Just as crossing the Rubicone marked a point of irrevocable action backed by collective intelligence, RUBICONe ensures that lane change decisions are made with comprehensive situational awareness and distributed consensus, showcasing the reliability gain of consensus in wireless communications. \end{abstract}


\section{Introduction}

Contemporary mission-critical communications are mainly based on wired infrastructure, where physical connections provide deterministic performance characteristics and established reliability metrics. Hardwired networks, particularly fiber-optic and high-grade copper interconnects, serve as the backbone for critical operations, offering predictable latency~\cite{9170905} (typically $\le 1$~ms), guaranteed bandwidth~\cite{8457339} and inherent security through physical isolation~\cite{6550872}. However, the emerging paradigm of mobility-centric applications, particularly in autonomous vehicles~\cite{parekh2022review} and industrial automation~\cite{frotzscher2014requirements}, requires the exploration of wireless alternatives, despite their intrinsic challenges in matching the deterministic nature of wired solutions~\cite{Xu2023WDC}.

The transition from wired to wireless domains in mission-critical scenarios represents a fundamental shift in reliability assurance mechanisms by overcoming the uncertainty of wireless connectivity. Although wired networks benefit from a stable physical layer and proven protocols, wireless communications introduce stochastic channel variations and environmental dependencies that challenge traditional reliability assumptions~\cite{Xu2020Raft}. Contemporary wireless solutions typically achieve reliability rates of 99.9\% (or 1000 hours of Mean Time Between Failures)~\cite{9356515} under optimal conditions, falling short of the 99.9999\% reliability standard commonly required in mission-critical applications~\cite{Xu2023WDC}. These constraints become particularly pronounced in applications demanding real-time consensus establishment, such as inter-vehicular communications, where maintaining wired-equivalent reliability becomes paramount for operational safety~\cite{8674597}.

While distributed consensus mitigates the single-point failure of individual perception through cooperative awareness, adapting rigid wired protocols like RAFT to lossy wireless ad-hoc networks remains a significant challenge due to packet loss and latency instability.

This paper presents RUBICONe (RAFT-Unified Behaviors for Inter-vehicular Cooperative Operations and Negotiations), a novel consensus-based architecture that aims to bridge the reliability gap between conventional wired infrastructure and emerging wireless solutions. By leveraging the open-source GNURadio platform to implement the IEEE 802.11p protocol, we construct a comprehensive wireless communication testbed capable of executing the complete fusion consensus protocol workflow. Our approach is guided by three key objectives:

    \begin{itemize}
        \item Implementation of a wireless RAFT consensus system using SDR devices to emulate realistic stochastic vehicular channels.
        \item Performance evaluation across a six-node topology to validate and calibrate the theoretical reliability models of RUBICONe.
        \item Quantitative assessment of system robustness against channel degradation to verify protocol stability in challenging wireless environments.
    \end{itemize}

By achieving these objectives, this study contributes to the field of wireless distributed consensus by enabling safer and more efficient lane change systems using RUBICONe with cost-effective hardware. To support open science and reproducibility, the complete implementation and experimental scripts have been made publicly available online\footnote{\url{https://anonymous.4open.science/r/V2V-Raft-SDR}}.

\section{System Model}
The RUBICONe architecture employs a RAFT-unified consensus algorithm tailored for vehicular networking scenarios to enable reliable lane-change decisions in autonomous vehicle networks. By adopting a RAFT-like coordination strategy, the system allows the acting node and other distributed vehicle nodes to collaboratively manage state transitions through dynamically adaptive RAFT roles (i.e., follower, candidate, leader), ensuring system resilience against unreliable links, dynamic topologies, and latency constraints inherent in open vehicular environments.

\begin{figure}[h]
    \centering
    \includegraphics[width=1\linewidth]{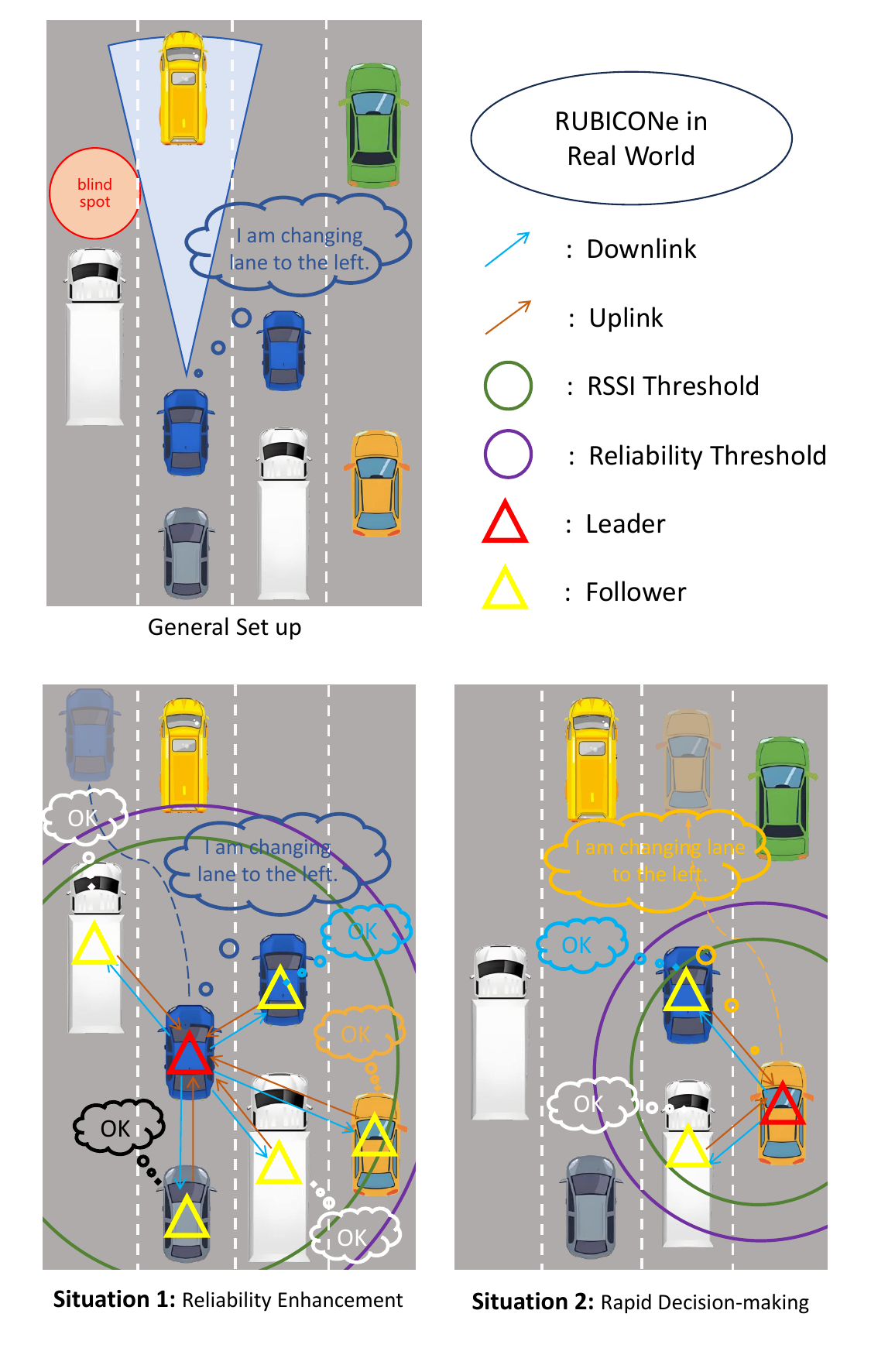}
    \caption{Lane-change decisions in RUBICONe.}
    \label{system model}
\end{figure}

As depicted in Fig. \ref{system model}, any vehicle (defined as the Subject Vehicle, SV) preparing to change lanes must first request permission from the cluster. Blind spots and occlusion often prevent the SV from detecting potential hazards alone. To address this, RUBICONe separates the role of the \textit{requester} from the \textit{decision-maker}.

RUBICONe derives collision-averse lane-change decisions by synthesizing voting inputs from distributed vehicular nodes. In this architecture, the elected leader acts as the central coordinator. For instance, in Situation 1 in Fig. \ref{system model}, a follower node (SV) transmits a lane-change request to the Leader. The Leader then aggregates multi-node consensus via downlink (DL) and uplink (UL) channels to validate the safety of the maneuver. Concurrently, as demonstrated in Situation 2, the Leader dynamically filters proximal neighbors through RSSI-weighted adaptive coordination, prioritizing high-signal nodes to expedite consensus for the SV's request.

\section{RUBICONe: The Algorithm}
The RUBICONe framework encompasses the wireless RAFT decision-making algorithms and their hardware implementation using low-cost devices.
This architecture consists of the following key components:

\subsection{System Organization}
The network comprises $N$ distributed nodes, each representing a vehicle within a vehicular ad-hoc network. Each node is implemented using an independent SDR device, enabling real-time consensus operations and local state management through programmable wireless physical-layer processing. The system's reliability---manifested as the correctness rate of consensus decisions---is fundamentally ensured by the majority-voting mechanism of the underlying protocol, which tolerates a bounded number of faulty or inconsistent inputs. 

In the experimental setup, an initial leader is elected autonomously by the protocol to bootstrap cluster formation.

\subsection{Fusion Consensus Protocol}

The inter-vehicular communication operates over IEEE 802.11p-compliant wireless channels, a protocol specifically designed for high-mobility vehicular environments with support for rapid topology changes and low-latency communication. These channels are characterized by time-varying characteristics including Rayleigh fading and Doppler shifts, implementing CSMA/CA for medium access control. This wireless environment introduces inherent challenges including packet loss, variable latency, and signal degradation---factors that the RUBICONe framework specifically addresses through its fusion consensus algorithm tailored for vehicular networking scenarios.

The consensus protocol functions as a distributed coordination mechanism where each vehicle represents an independent node in the system, communicating via onboard wireless links. Vehicles periodically broadcast their perceptual information and initiate or participate in consensus-based decision processes when preparing to change lanes, ensuring that irreversible actions such as lane changes are executed with higher reliability and consistency at the group level. The complete workflow consists of the following three phases:

\begin{enumerate}
    \item \textbf{System Initialization}
    
    During initialization, each vehicle node begins periodic broadcasting of heartbeat messages to discover neighboring participants and establish a coherent network view. These heartbeat messages contain: node ID, current position and velocity, average RSSI over the last $K$ intervals, signal-to-noise ratio, and a timestamp. Each node maintains a set of locally tracked metrics, including average RSSI, SNR, channel busy ratio (CBR), and a dynamic neighbor list \(\mathcal{N}_i\).

    This neighbor information and the associated link metrics undergo continuous updates and synchronization throughout the network, ensuring the maintenance of system coherence and enabling rapid detection of topology changes.
    
    \item \textbf{Leader Election}
    
    The leader election process ensures fast convergence through a streamlined multi-phase protocol:

\begin{itemize}
    \item \textbf{Broadcast Phase:} Nodes initialize as candidates and periodically broadcast heartbeat messages containing their state to announce candidacy.
    
    \item \textbf{Collection Phase:} Nodes collect broadcasts and independently score candidates based primarily on SNR. The candidate with the highest link quality is identified as the preferred leader.
    
    \item \textbf{Announcement Phase:} The preferred candidate broadcasts a \textit{Leader Request} containing log freshness indicators (Term/Index) and a digital signature for security validation.
    
    \item \textbf{Verification Phase:} Receivers validate the request's integrity. Upon success, they broadcast an ACK; otherwise, a NACK is sent with logged rejection reasons.
    
    \item \textbf{Establishment Phase:} The leader broadcasts a confirmation upon receiving a quorum of ACKs. If the quorum is not met, the node reverts to follower status to trigger re-election.
\end{itemize}

The election timeout threshold is dynamically calculated based on real-time quality indicators \(\gamma_{ij}\) and neighbor density:
    
    \begin{equation}
    T_{\text{election}, i} = \left( 1 + \frac{\alpha}{\sum_{j \in \mathcal{N}_i} \gamma_{ij}} \right) T_{\text{base}}\,.
\end{equation}
    
    Here, \(T_{\text{base}}\) denotes the baseline timeout duration, which is a configurable system parameter, and \(\alpha\) is a scaling factor that adjusts the sensitivity to channel quality variations.
    
    \item \textbf{Lane Change Proposal and Consensus}
    
    The decision process follows a Client-Leader interaction model tailored for vehicular safety:
    
    \begin{itemize}
        \item \textbf{Request Initiation:} When a Subject Vehicle (SV) intends to change lanes, it first performs a local safety check. If the local check passes, the SV sends a \textit{Log Entry Proposal} (containing the intended maneuver and timestamp) to the current Leader.
        
        \item \textbf{Proposal Scheduling:} The Leader collects requests from all SVs. To prevent conflicting maneuvers, the Leader prioritizes requests based on a safety-critical score:
        \begin{equation}
            \text{Priority}_{\text{req}} = \frac{|\Delta \mathbf{v}_{\text{req}}|}{d_{\text{req}}} + \frac{\lambda}{\gamma_{\text{req}}}\,,
        \end{equation}
        where requests from vehicles with higher relative velocity or closer proximity are processed first. The Leader then appends the highest-priority request to its log and broadcasts an \textit{AppendEntries} RPC to all followers to initiate voting.

        \item \textbf{Signal-Aware Tie-Breaking Strategy:}
Standard RAFT relies on randomized timeouts to resolve split votes, which introduces latency variance that is unacceptable for high-mobility vehicular contexts. To expedite consensus and guarantee system liveness, RUBICONe introduces a deterministic tie-breaking mechanism.
While strictly adhering to the majority rule for clear consensus, nodes with higher SNR are assigned marginally higher weights to resolve even-split scenarios deterministically.

Upon receiving the leader's proposal, followers validate it against their local perception. The Leader calculates the dynamic weight $w_i$ for each voting node $i$:
\begin{equation}
    w_i = 1 + \varepsilon \cdot \frac{\gamma_i - \gamma_{\min}}{\Delta \gamma}\,,
\end{equation}
where $\gamma_{\min}$ and $\Delta \gamma$ are the minimum SNR and SNR range of the current voting batch. The bias factor $\varepsilon = 10^{-3}$ is set deliberately small to ensure that $w_i$ never overrides a legitimate integer majority (i.e., $3$ votes vs. $2$ votes), functioning solely as a tie-breaker when vote counts are equal.
        
        The Leader commits the decision and notifies the SV only if the weighted approval exceeds the weighted rejection:
        \begin{equation}
            \sum_{i \in \mathcal{V}_{\text{yes}}} w_i > \sum_{j \in \mathcal{V}_{\text{no}}} w_j\,.
        \end{equation}
        
        This mechanism employs channel quality as a cross-layer heuristic for proximity and perception reliability, resolving split-vote stalemates without triggering time-consuming leader re-elections. By enforcing a decisive outcome based on link quality, RUBICONe prioritizes decision availability over the latency of re-election, ensuring the system never hangs in an indeterminate state during critical maneuvers.
    \end{itemize}

\end{enumerate}

\subsection{Evaluation Model}
To comprehensively evaluate the performance of RUBICONe in dynamic vehicular networks, an evaluation model is proposed, focusing on two core aspects: system reliability and robustness. For the evaluation, we assume the cluster membership remains stable during the short duration of a single negotiation epoch ($<$100 ms), with dynamic topology changes handled between epochs.

\begin{itemize}
    \item \textbf{System Reliability (\(P_{\text{sys}}\)):} System reliability refers to the probability that the distributed vehicle network reaches a correct consensus decision.
Given that the weighting factor $\varepsilon$ is negligible ($\varepsilon \ll 1$) and does not alter the outcome of non-tie scenarios, we approximate the reliability using a standard majority voting model based on the Binomial distribution.
The reliability is computed as follows:

\begin{equation}
\begin{aligned}
    P_{\text{sys}} = & \sum_{k=\lfloor \frac{N}{2} \rfloor + 1}^{N} \binom{N}{k} p_{\text{node}}^k (1-p_{\text{node}})^{N-k} \\[1ex]
              & + \frac{1}{2} \cdot \mathbb{I}_{\{N \text{ is even}\}} \cdot \binom{N}{N/2} p_{\text{node}}^{N/2} (1-p_{\text{node}})^{N/2}\,,
\end{aligned}
\label{eq:reliability}
\end{equation}

    where \(N\) represents the total number of participating nodes, \(p_{\text{node}}\) denotes the individual node reliability (the probability that a node provides accurate perception data), \(k\) is the number of nodes voting in favor, and \(\mathbb{I}_{\{N \text{ is even}\}}\) is an indicator function that equals 1 when \(N\) is even and 0 otherwise. The first term calculates the probability that more than half of the nodes agree, while the second term handles the edge case of an even split. In the practical RUBICONe implementation, strict even splits are resolved deterministically using the SNR-weighted method described in Sec. III-B, ensuring that the decision aligns with the higher-quality links. The term with coefficient $\frac{1}{2}$ in Eq. \eqref{eq:reliability} serves as a conservative baseline for even-split scenarios, representing the statistical expectation of a correct decision.
    \textit{Note: We assume independent packet loss for the baseline model, while experimental results in Sec. IV implicitly capture real-world correlated channel effects.}

    \item \textbf{Robustness (\(\eta_r\)):} Robustness quantifies the system's ability to maintain performance levels under minor disturbances or environmental fluctuations. It is defined as an exponential decay function that compares the current system reliability to its baseline value:

    \begin{equation}
        \eta_r = e^{-\kappa\left(1 - \frac{P_{\text{current}}}{P_{\text{baseline}}}\right)}
        \label{eq:robustness}\,,
    \end{equation}

    where \(P_{\text{current}}\) represents the system reliability under disturbed conditions, \(P_{\text{baseline}}\) denotes the reliability under nominal operating conditions, and \(\kappa > 0\) is the interference sensitivity coefficient that controls how rapidly robustness degrades as performance deviates from the baseline.
\end{itemize}

This study leverages insights from communication studies, emphasizing the role of consensus algorithms like RAFT in ensuring robust decision-making. By addressing current limitations, this paper lays the foundation for safer and more intelligent road traffic systems.

\section{Evaluations and Experiments}

This section employs experimental validation with hardware and numerical simulation to rigorously verify the model's effectiveness.

\subsection{Experimental Setup and Design}
\begin{figure}[ht]
    \centering
    \includegraphics[width=1.0\linewidth]{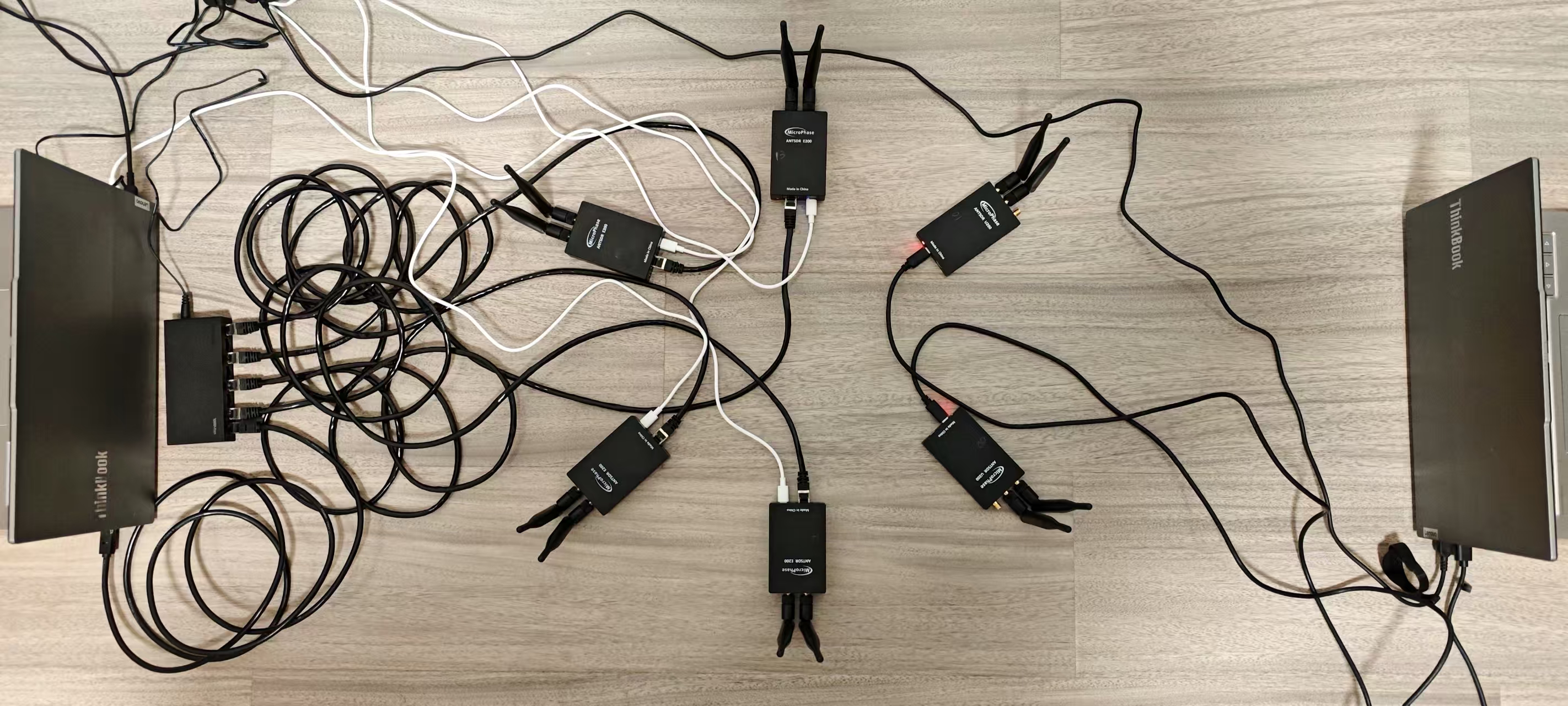}
    \caption{Experimental equipment of \textit{MicroPhase} ANTSDR devices.}
    \label{fig:shiwu}
\end{figure}

The experiment employs \textit{MicroPhase} ANTSDR devices as the primary node platform, as shown in Fig. \ref{fig:shiwu}. The use of both U200 and E200 models introduces hardware-level diversity, effectively emulating the randomness and heterogeneity inherent in real-world vehicular environments.
All inter-node communications are built upon a complete IEEE 802.11p protocol stack implemented using the open-source GNURadio platform, which provides the foundational physical and link-layer toolchain for vehicular ad-hoc networking. The fusion consensus protocol is developed and executed atop this reliable wireless communication framework.

This study integrates hardware-in-the-loop experiments with simulation-based performance analysis using a testbed of \textit{MicroPhase} ANTSDR nodes. 

The experimental configuration consists of a six-node cluster executing the integrated consensus protocol. By systematically varying the SNR during protocol operation, we aim to observe the following key relationships:

\begin{enumerate}
    \item \textbf{System Scale vs. SNR}: Analysis of how system scalability interacts with channel conditions.
    
    \item \textbf{System Reliability vs. Node Count}: Investigation of the relationship between overall system trustworthiness and the number of participating nodes, assuming fixed individual node reliability. Two specific SNR scenarios corresponding to packet loss rates of 30\% and 70\% are selected. For each scenario, multiple individual node reliability levels are configured to measure the variation in overall system reliability.
    
    \item \textbf{System Reliability vs. Individual Node Reliability}: Examination of how system reliability responds to changes in individual node reliability, with a fixed node count. This analysis considers the combined effects of SNR variation (which indirectly affects cluster fault tolerance through packet loss) and dynamic individual node reliability (which directly determines system reliability).
\end{enumerate}

Table \ref{tab:params} summarizes the key parameters used throughout our experiment.

\begin{table}[ht]
    \centering
    \caption{System Parameters of Testbed.}
    \label{tab:params}
    \begin{tabular}{ll}
    \toprule
    \textbf{Parameter} & \textbf{Value} \\
    \midrule
    Center Frequency & 5.9 GHz\\
    Channel Bandwidth & 10 MHz \\
    Clock Accuracy & $\pm$0.5 ppm VCTCXO \\
    Noise Figure & $<$8 dB \\
    Data Rate & 3–27 Mbit/s \\
    \bottomrule
    \end{tabular}
\end{table}

\subsection{Result Analysis}

\subsubsection{System Scale vs. SNR}

\begin{figure}[ht]
    \centering
    \includegraphics[width=1.0\linewidth]{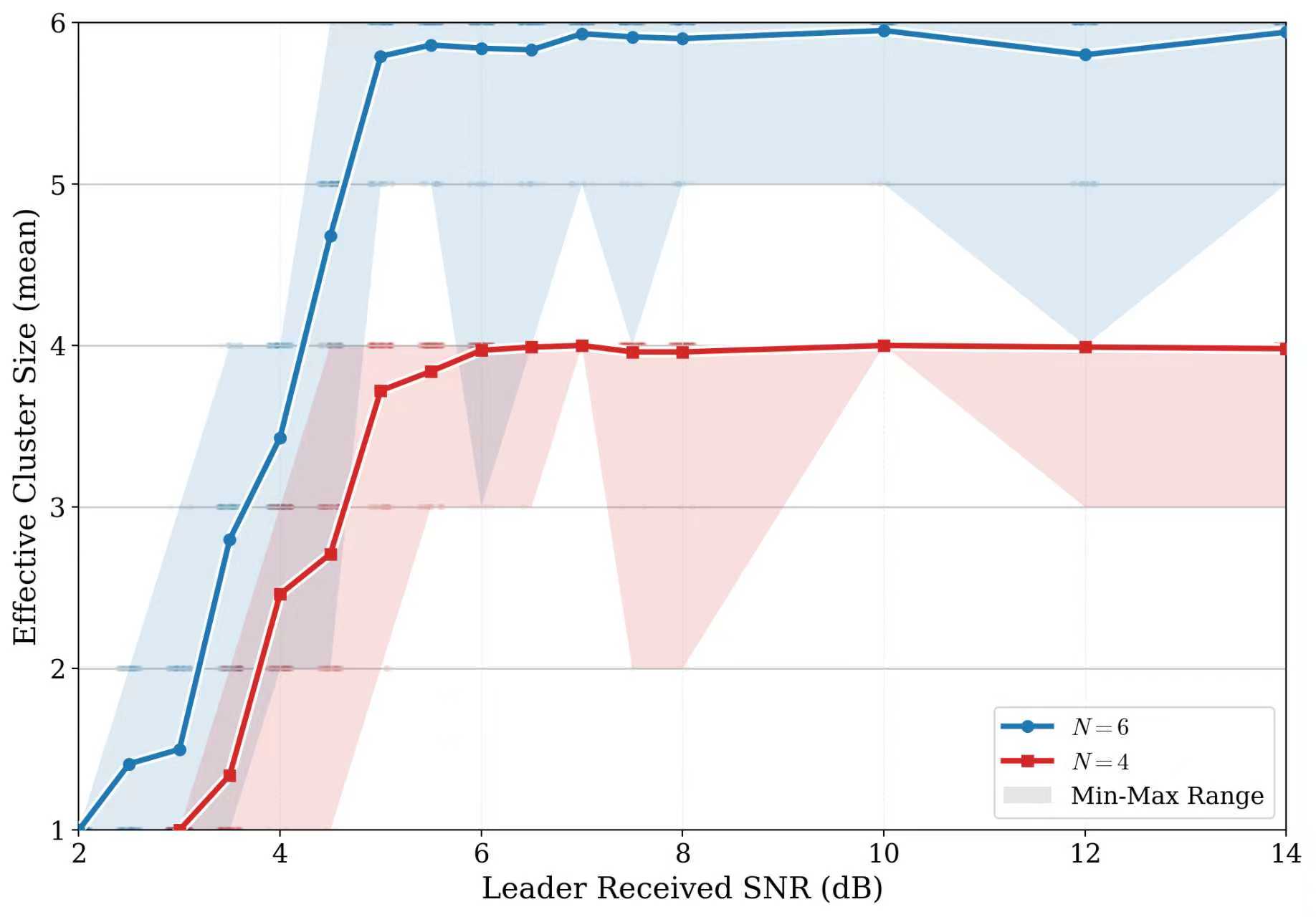}
    \caption{Variation of effective cluster size with leader received SNR under different initial node counts ($N=4$ and $N=6$).}
    \label{fig:exp1}
\end{figure}

To evaluate the robustness of RUBICONe under realistic vehicular channel conditions, we conducted experiments measuring the relationship between the effective node cluster size and the leader received SNR. As depicted in Fig.~\ref{fig:exp1}, the results illustrate how wireless channel quality directly impacts the connectivity and scale of the distributed consensus cluster.

Analysis of the results reveals two key trends. First, cluster connectivity correlates directly with channel quality; as SNR degrades, packet loss triggers node disconnection, though stability is maintained above 8 dB. Second, while larger clusters offer higher theoretical fault tolerance, they exhibit greater sensitivity to channel degradation. As shown in Fig.~\ref{fig:exp1}, the $N=6$ cluster experiences a sharper reduction in effective size at low SNR than $N=4$, as higher node density exacerbates collision probabilities under multi-access interference.

\subsubsection{System Reliability vs. Node Scale and Individual Node Reliability}

\begin{figure}[ht]
    \centering
    \includegraphics[width=0.95\linewidth]{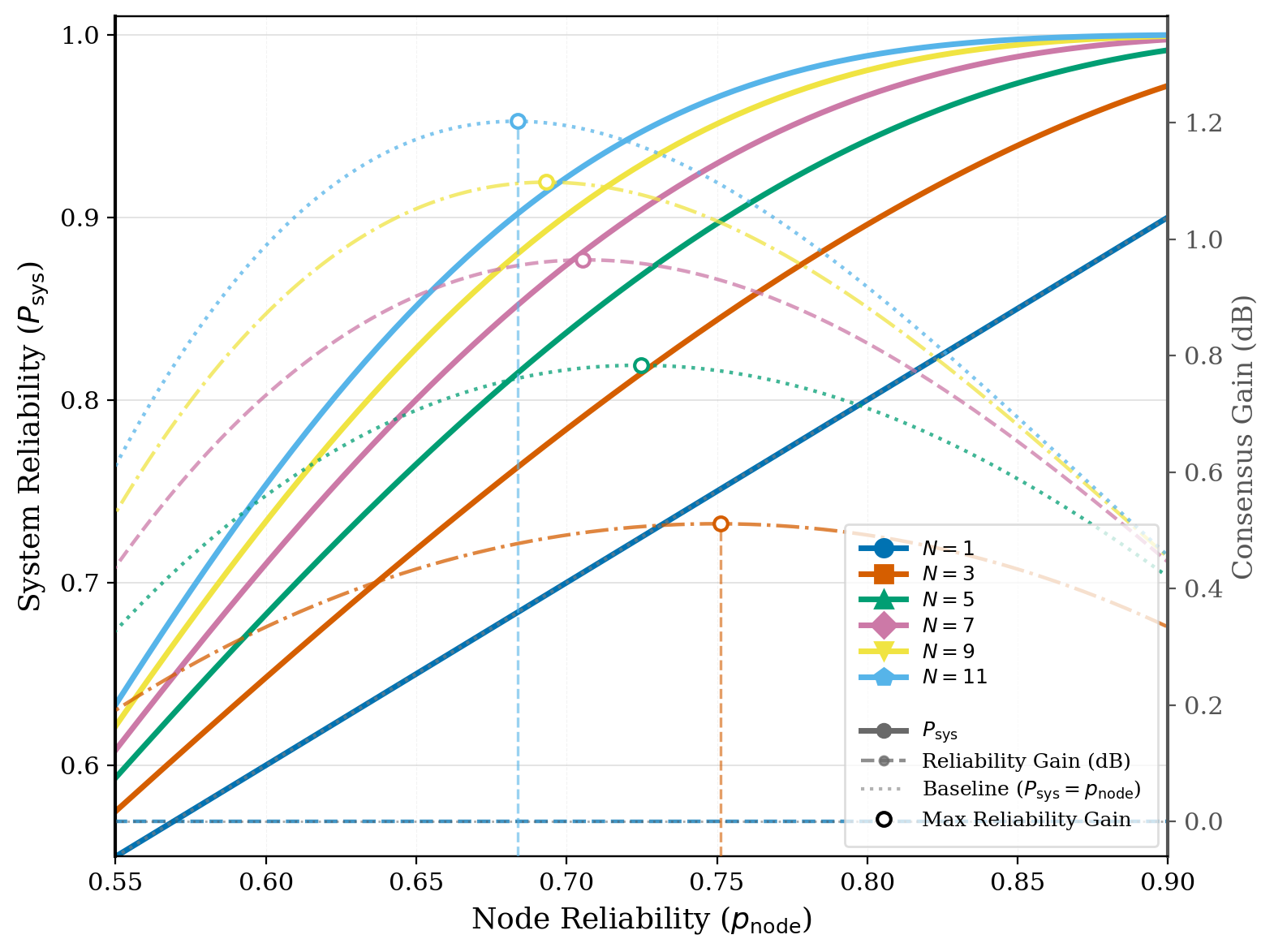}
    \caption{$P_{\text{sys}}$ as a function of $p_{\text{node}}$ for different cluster sizes ($N$ = 1, 3, 5, 7, 9, 11), simulated using the consensus reliability model.}
    \label{fig:simulation}
\end{figure}

\begin{figure}[ht]
    \centering
    \includegraphics[width=0.95\linewidth]{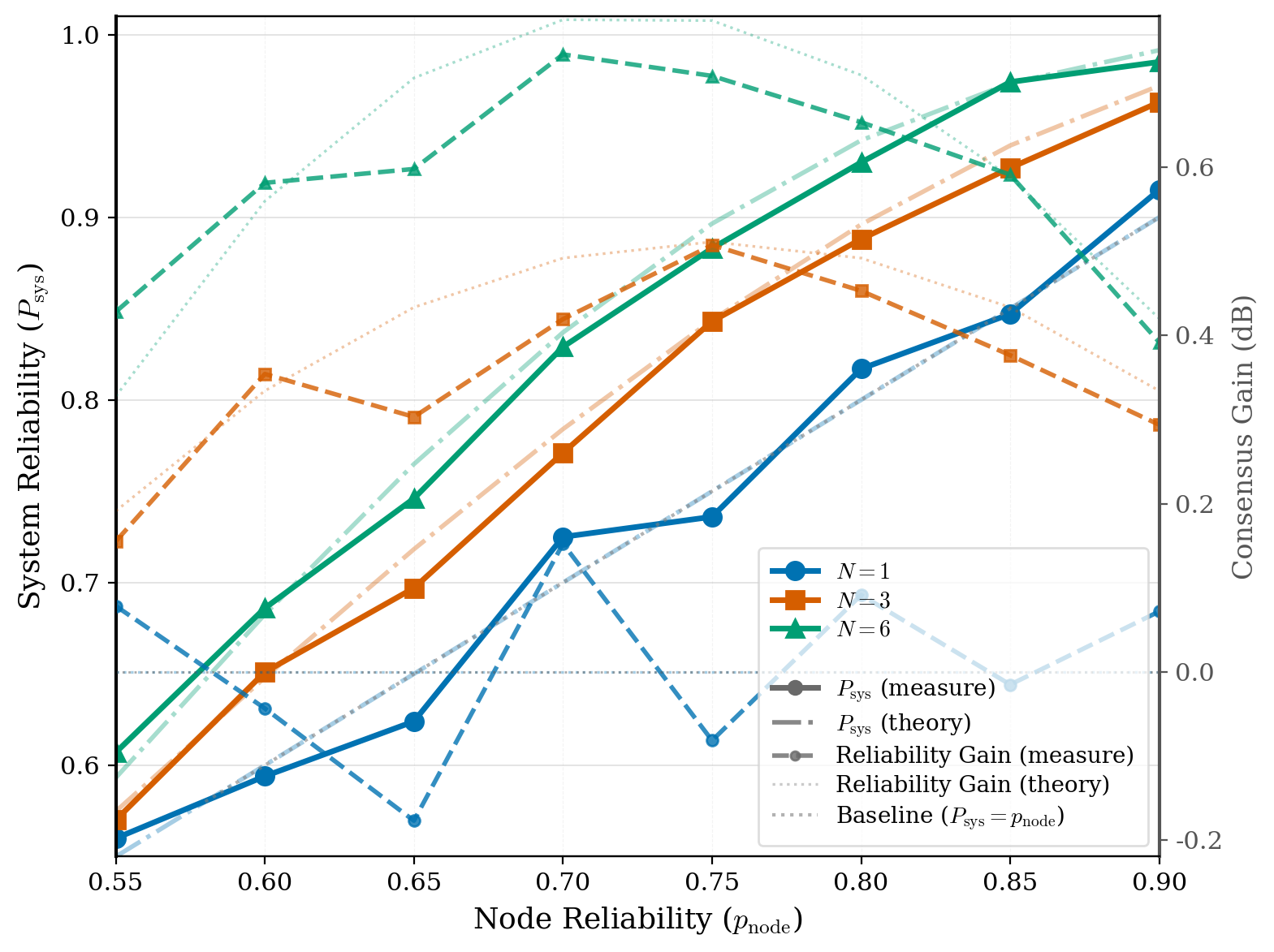}
    \caption{Measured and theoretical system reliability under high-quality channel conditions (SNR = 14 dB) for cluster sizes $N$ = 1, 3, and 6.}
    \label{fig:exp_14db}
\end{figure}

\begin{figure}[ht]
    \centering
    \includegraphics[width=0.95\linewidth]{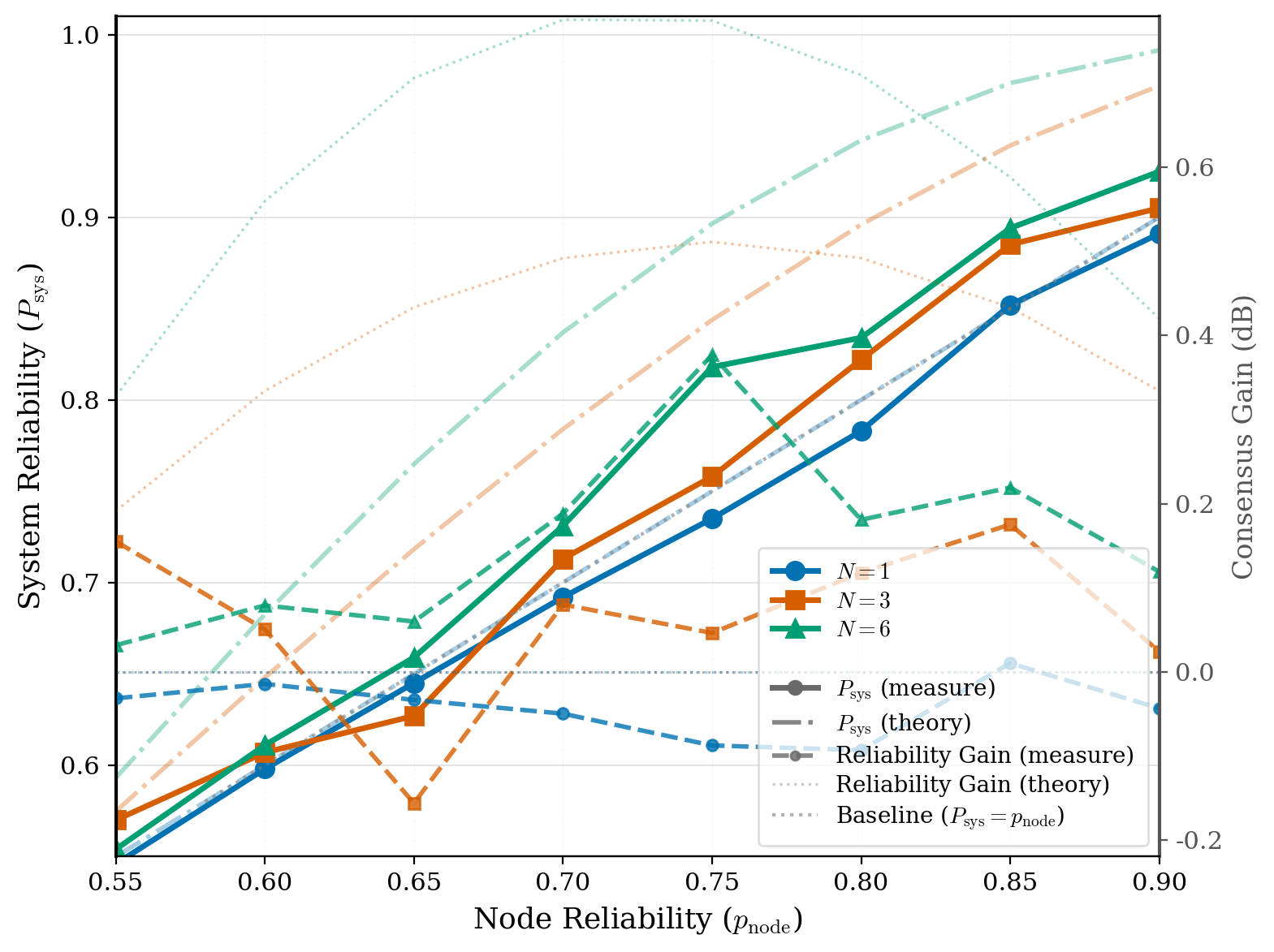}
    \caption{Measured and theoretical system reliability under challenging channel conditions (SNR = 4 dB) for cluster sizes $N$ = 1, 3, and 6.}
    \label{fig:exp_4db}
\end{figure}

To characterize the fundamental relationships between system reliability ($P_{\text{sys}}$), cluster size ($N$), and individual node reliability ($p_{\text{node}}$), we conducted integrated software simulations and hardware experiments. The simulation employed the system reliability model presented in the Evaluation Model section, while hardware experiments were performed on the SDR testbed under two controlled SNR conditions (4 dB and 14 dB). The combined results, illustrated in Figs. \ref{fig:simulation}, \ref{fig:exp_14db}, and \ref{fig:exp_4db}, quantify these key relationships and validate the theoretical model under real-world channel effects.

The analysis of simulation and experimental data reveals several key insights:

\begin{enumerate}
    \item \textbf{System reliability increases with node count but exhibits diminishing returns.} For a fixed $p_{\text{node}}$, both simulation (Fig.~\ref{fig:simulation}) and experimental results (Figs. \ref{fig:exp_14db} and \ref{fig:exp_4db}) confirm that $P_{\text{sys}}$ rises as the cluster size $N$ increases. This demonstrates the fundamental advantage of collective decision-making, where leveraging information from multiple vehicles significantly enhances the accuracy of lane-change feasibility assessment. However, the rate of reliability gain progressively diminishes. As shown in the simulation, the marginal improvement from adding nodes becomes minimal beyond a certain scale (e.g., from $N=5$ to $N=11$). This indicates that after reaching a sufficient voting quorum, introducing additional nodes provides limited extra benefit to overall trustworthiness, guiding practical system design toward an optimal size.

    \item \textbf{The optimal system scale depends on the required reliability level and channel conditions.} The impact of cluster size on $P_{\text{sys}}$ varies under different $p_{\text{node}}$ and SNR conditions. Under high SNR (14 dB, Fig.~\ref{fig:exp_14db}), a larger cluster consistently provides higher $P_{\text{sys}}$ for a given $p_{\text{node}}$. In contrast, under low SNR (4 dB, Fig.~\ref{fig:exp_4db}), the benefit of scaling is attenuated due to increased packet loss and link instability. Furthermore, as indicated by the simulation (Fig.~\ref{fig:simulation}), when $p_{\text{node}}$ is already high (e.g., $>$0.8), a small cluster may achieve satisfactory reliability, whereas a larger cluster is necessary to compensate for moderate individual node reliability. This confirms that the optimal node count is scenario-dependent, varying with both node-level performance and channel quality.

    \item \textbf{Experimental validation of the theoretical reliability model.} As shown in Fig. \ref{fig:exp_14db}, the measured $P_{\text{sys}}$ under high SNR conditions closely aligns with the theoretical predictions. This validates that enhancing $p_{\text{node}}$ translates directly to system-level gains without significant implementation overhead. For instance, under high SNR (14 dB, Fig.~\ref{fig:exp_14db}), a 6-node system shows a measurable gain in $P_{\text{sys}}$ as $p_{\text{node}}$ increases from 0.70 to 0.90. This underscores the importance of improving each vehicle's onboard sensing and processing capabilities as a fundamental means to boost overall system reliability.

\end{enumerate}

In summary, this analysis elucidates the key relationships governing the reliability of the RUBICONe consensus framework: $P_{\text{sys}}$ scales positively but sub-linearly with $N$, is positively correlated with $p_{\text{node}}$, and is moderated by channel quality. These findings provide a quantitative foundation for designing reliable vehicular networks, balancing the benefits of collective intelligence against the costs of scale and the constraints of the wireless medium.

\subsection{Discussion}
In hardware experiments, reliability fluctuated around theoretical predictions due to the inherent stochasticity of real-world wireless environments, rather than following a perfectly monotonic trend. These variations primarily stem from:
\begin{itemize}
    \item \textbf{Time-Varying Wireless Channel Dynamics:} Even under a controlled average SNR, the instantaneous channel condition is subject to fast fading, multi-path propagation, and Doppler shifts caused by relative movement. This leads to bursty packet loss and fluctuating link quality, causing trial-to-trial variations in consensus message delivery times and success rates.
    \item \textbf{Non-Ideal Hardware Effects:} The use of software-defined radios, while flexible, introduces subtle impairments, which can affect the precise timing and decoding of packets, adding a layer of randomness not captured in discrete packet loss models.
\end{itemize}

\section{Conclusion}

This paper presents RUBICONe, a vehicular consensus framework implemented on SDR nodes to bridge the gap between wired protocols and stochastic wireless environments. Our experimental results demonstrate significant improvements in key metrics: the proposed signal-aware mechanism ensures superior system reliability and robustness in multi-node clusters, effectively mitigating the impact of channel instability.

Future work will expand the system's scope from crash-fault tolerance to security-critical scenarios. Specifically, we aim to incorporate Byzantine Fault Tolerance (BFT) mechanisms to defend against malicious or compromised nodes. Additionally, further research will focus on dynamic consensus optimization under high-mobility channel states and rigorous latency profiling, investigating adaptive algorithms to minimize convergence time while maintaining real-time state consistency.

	\bibliography{ref}

@ARTICLE{Xu2020Raft,
  author={Xu, Hao and Zhang, Lei and Liu, Yinuo and Cao, Bin},
  journal={IEEE Wireless Communications Letters}, 
  title={RAFT Based Wireless Blockchain Networks in the Presence of Malicious Jamming}, 
  year={2020},
  volume={9},
  number={6},
  pages={817-821},
  doi={10.1109/LWC.2020.2971469}
}

@ARTICLE{Xu2023WDC,
  author={Xu, Hao and Fan, Yixuan and Li, Wenyu and Zhang, Lei},
  journal={IEEE Internet of Things Journal}, 
  title={Wireless Distributed Consensus for Connected Autonomous Systems}, 
  year={2023},
  volume={10},
  number={9},
  pages={7786-7799},
  doi={10.1109/JIOT.2022.3229746}
}

@ARTICLE{9170905,
  author={Lu, Yunlong and Huang, Xiaohong and Zhang, Ke and Maharjan, Sabita and Zhang, Yan},
  journal={IEEE Transactions on Industrial Informatics}, 
  title={Low-Latency Federated Learning and Blockchain for Edge Association in Digital Twin Empowered 6G Networks}, 
  year={2021},
  volume={17},
  number={7},
  pages={5098-5107},
  doi={10.1109/TII.2020.3017668}
}

@INPROCEEDINGS{8457339,
  author={Moravejosharieh, Amir Hossein and Watts, Michael J. and Song, Yu},
  booktitle={2018 15th International Joint Conference on Computer Science and Software Engineering (JCSSE)}, 
  title={Bandwidth Reservation Approach to Improve Quality of Service in Software-Defined Networking: A Performance Analysis}, 
  year={2018},
  volume={},
  number={},
  pages={1-6},
  doi={10.1109/JCSSE.2018.8457339}
}

@ARTICLE{6550872,
  author={Adeli, Majid and Liu, Huaping},
  journal={IEEE Communications Letters}, 
  title={On the Inherent Security of Linear Network Coding}, 
  year={2013},
  volume={17},
  number={8},
  pages={1668-1671},
  doi={10.1109/LCOMM.2013.062113.130478}
}

@ARTICLE{8674597,
  author={Zhou, Jianshan and Tian, Daxin and Wang, Yunpeng and Sheng, Zhengguo and Duan, Xuting and Leung, Victor C.M.},
  journal={IEEE Transactions on Mobile Computing}, 
  title={Reliability-Optimal Cooperative Communication and Computing in Connected Vehicle Systems}, 
  year={2020},
  volume={19},
  number={5},
  pages={1216-1232},
  doi={10.1109/TMC.2019.2907491}
}

@article{parekh2022review,
  title={A review on autonomous vehicles: Progress, methods and challenges},
  author={Parekh, Darsh and Poddar, Nishi and Rajpurkar, Aakash and Chahal, Manisha and Kumar, Neeraj and Joshi, Gyanendra Prasad and Cho, Woong},
  journal={Electronics},
  volume={11},
  number={14},
  pages={2162},
  year={2022},
  publisher={MDPI}
}

@inproceedings{frotzscher2014requirements,
  title={Requirements and current solutions of wireless communication in industrial automation},
  author={Frotzscher, Andreas and Wetzker, Ulf and Bauer, Matthias and Rentschler, Markus and Beyer, Matthias and Elspass, Stefan and Klessig, Henrik},
  booktitle={2014 IEEE international conference on communications workshops (ICC)},
  pages={67--72},
  year={2014},
  organization={IEEE}
}

@ARTICLE{9356515,
  author={Herlich, Matthias and Maier, Christian},
  journal={IEEE Communications Magazine}, 
  title={Measuring and Monitoring Reliability of Wireless Networks}, 
  year={2021},
  volume={59},
  number={1},
  pages={76-81},
  doi={10.1109/MCOM.001.2000250}
}
	\bibliographystyle{IEEEtran}
\end{document}